\documentclass[12pt]{article}

\begin{document}
\centerline{\Large\bf Past and Future of Cosmic Topology}
\vskip 1cm
\centerline{Jean-Pierre Luminet}
\centerline{Observatoire de Paris-Meudon,}
\centerline{D\'epartement d'Astrophysique Relativiste et de 
Cosmologie,} 
\centerline{CNRS UPR-176, F-92195 Meudon Cedex, 
France} 
\centerline{luminet@obspm.fr}
\vskip 1cm

\centerline{\Large\bf Abstract}
\vskip 1cm 

The global topology of the universe, including questions about the 
shape of space, its volume and its connectedness, is a fundamental 
issue in cosmology which has been overlooked for many years,
except by some pioneering authors (see references 
in Lachi\`eze-Rey \& Luminet, 1995, hereafter LaLu95). In the first 
part of the present article, I set out some unexplored historical material 
about the early development of cosmic topology. It stems out that the two ``fathers" of 
the big bang concept, Friedmann and Lema\^{\i}tre, were also the first 
to realize the full importance of cosmic topology, whereas Einstein 
remained reluctant to the idea of a multi--connected space. In the 
second part I briefly  comment new developments in the field since 1995, both from 
a theoretical and an observational point of view. They fully confirm 
that cosmic topology is, more than ever, a promising field of 
investigation.

\section {Birth of cosmic topology}

One of the oldest cosmological questions is the physical extension of 
space : is it finite or infinite?  (see e.g.  Luminet, 1994 ; Luminet 
\& Lachi\`eze-Rey, 1994). In the history of cosmology, it is well 
known that the Newtonian physical space, 
mathematically identified with infinite Euclidean space ${\bf R}^3$, gave 
rise to paradoxes such as darkness of night (see e.g.  
Harrison, 1987) and to problems of boundary conditions. Regarding for instance the Mach's 
idea according to which 
local inertia would result from the contributions of masses at 
infinity, an obvious divergence difficulty arose, since a  
homogeneous Newtonian universe with non--zero density had an infinite mass.

The aim of relativistic cosmology was to deduce 
from gravitational field equations some physical models of 
the universe as a whole. When Einstein (1917) assumed in his static cosmological 
solution that space was a positively--curved hypersphere, one of 
his strongest motivations was to provide a model for a {\it finite} space, 
although without a boundary. He regarded the closure of space as 
necessary to solve the problem of inertia (Einstein, 1934). 
The spherical model cleared up most of the paradoxes 
stemmed from Newtonian cosmology in such an elegant way 
that most cosmologists of the time adopted the new paradigm of a 
closed space, without 
examining other geometrical possibilities. Einstein was also 
convinced that the hypershere provided not only the metric of cosmic 
space -- namely its local geometrical properties -- but also its 
global structure, namely its topology.  
However, topology does not seem to have been a major preoccupation of Einstein; 
 his 1917 cosmological article did not mention any topological 
alternative to the spherical space model.

 Some of his colleagues pointed out to 
Einstein the arbitrariness of his choice. The reason was the 
following. The  global shape of space is not only depending 
on the metric; it primarily depends on its topology, and requires a 
complemenraty approach to Riemannian differential geometry.  Since Einstein's 
equations are partial derivative equations, they describe only local 
geometrical properties of spacetime. The latter are contained in the 
metric tensor, which enables us to calculate the components of 
the curvature tensor at any non-singular point of spacetime.  But 
Einstein's equations do not fix the global structure of spacetime : to 
a given metric solution of the field equations, correspond several 
(and in most cases an infinite number of) topologically distinct 
universe models. 

First, De Sitter (1917) noticed that the Einstein's 
solution admitted a different spaceform, namely the 3-dimensional 
projective space (also called elliptic space), obtained from the hypersphere 
by identifying antipodal points.  The projective 
space has the same metric as the spherical one, but a different 
topology (for instance, for the same curvature radius its volume is twice 
smaller).  

Next  H. Weyl pointed out the freedom of choice between  spherical and 
elliptical topologies.  The Einstein's answer  (1918) was unequivocal : ``Nevertheless 
I have like an obscure feeling which leads me to prefer the spherical model.
 I have the presentiment 
that manifolds in which any closed curve can be continuously 
contracted to a point are the simplest ones.  Other persons must 
share this feeling, otherwise astronomers would have taken into 
account the case where our space is Euclidean and finite.  
Then the two-dimensional Euclidean space would have the connectivity 
properties of a ring surface. It is an Euclidean plane in which any 
phenomenon is doubly periodic, where points located in the same 
periodical grid are identical. In finite Euclidean space, three 
classes of non continuously contractible loops would exist.  In a similar  
 way, the elliptical space possesses a class of non continuously contractible loops,
 contrary to the spherical case; it is the reason 
why I like it less than the spherical space.  Can it be proved that 
elliptical space is the only variant of spherical space?  It seems yes 
to me". 

 Einstein (1919)  repeated his argumentation in a postcard 
 sent to Felix Klein : ``I would like to give you a reason why the spherical
 case should be preferred to 
the elliptical case.  In spherical space, any closed curve can be 
continuously contracted to a point, but not in the elliptical space; 
in other words the spherical space alone is simply-connected, not the 
elliptical one [...]  Finite spaces of 
arbitrary volume with the Euclidean metric element undoubtedly exist, which can be 
obtained from infinite spaces by assuming a triple periodicity, 
namely identity between sets of points.  However such possibilities, 
which are not taken into account by general relativity, have the wrong 
property to be multiply-connected".  From these remarks it follows that the
 Einstein's  prejudice in favour of 
simple--connectedness of space was of an aesthetical nature, rather than 
being based on physical reasoning.

  In his answer to Weyl, Einstein was 
definitely wrong on the last point : in dimension three,  an infinite 
number of topological variants of the spherical space -- all closed -- do exist,  
including the so-called lens spaces (whereas in dimension 
two, only two spherical spaceforms exist, the ordinary sphere and 
the elliptic plane).  However, nobody 
knew this result in the 1920's : the topological classification of 
3--dimensional spaces was still at its beginnings.  The study of 
 Euclidean spaceforms started in the context of 
crystallography.  Feodoroff (1885) classified the 18 symmetry groups 
of crystalline structures in ${\bf R}^3$, Bieberbach (1911) developed 
a full theory of crystallographic groups, and  twenty years later only
Nowacki (1934) showed how the Bieberbach's results could be applied to 
achieve the 
classification of 3--dimensional Euclidean spaceforms.  The case of spherical spaceforms
 was first set by Klein (1890) and  
Killing (1891).  The problem was fully solved much later (Wolf, 1960).  
Eventually, the classification of homogeneous hyperbolic spaces was 
impulsed in the 1970's ; it is now an open field of intensive 
mathematical research (Thurston, 1979, 1997).

Going back to relativistic cosmology, the discovery of non-static 
solutions by Friedmann (1922) and, independently, Lema\^{\i}tre (1927),
 opened a new era for models of the universe as a whole (see, e.g., 
 Luminet, 1997 for an epistemological analysis).  
Although Friedmann and Lema\^{\i}tre are generally considered as the discoverers
 of the big bang  concept --at least of the notion of a dynamical universe 
evolving from an initial singularity --,  one of their most original considerations, devoted
 to the topology of space,  was 
overlooked.  As they stated, the 
homogeneous isotropic universe models (F--L models)  admit spherical, Euclidean or 
hyperbolic space sections according to the sign of their (constant) 
curvature (respectively positive, zero or negative).  
In addition, Friedmann (1923) pointed out the topological indeterminacy of the 
solutions in his popular book on general relativity, and he emphasized how the 
Einstein's theory was 
unable to deal with the global structure of spacetime.  He gave the 
simple example of the cylinder -- a  locally Euclidean surface which has not
 the same topology as the plane.  Generalizing the argument to  higher 
dimensions, he concluded that several topological spaces could  
be used to describe the same solution of Einstein's equations.  

Topological considerations were fully developed in his second 
cosmological article (Friedmann, 1924),
although primarily devoted to the analysis of  hyperbolic 
solutions. Friedmann clearly outlined  the 
fundamental limitations of relativistic cosmology :``Without 
additional assumptions, the Einstein's world equations do not answer 
the question of the finiteness of our universe", he wrote.  Then he 
described how space could become finite (and multi-connected) by 
suitably identifying  points.  He also predicted the possible 
existence of ``ghost" images of astronomical sources, since at the same 
point of a multi--connected space an object and its ghosts would coexist. 
 He  added  
that ``a space with positive curvature is always finite", but he 
recognized the fact that the mathematical knowledge of his time did not 
allow him to ``solve the question of finiteness for a negatively--curved  space". 
 
 Comparing with Einstein's reasoning, it appears that the 
Russian cosmologist had no prejudice in favour of a simply-connected 
topology.  Certainly Friedmann believed that only spaces with 
finite volume were 
physically realistic.  Prior to his discovery of hyperbolic solutions, the 
cosmological solutions derived by Einstein, de Sitter and himself 
had a positive spatial curvature, thus a finite volume.  With negatively--curved spaces,
the situation became problematic, because the 
``natural" topology of hyperbolic space has an infinite volume. It is 
the reason why Friedmann, in order to justify the physical pertinence 
of his solutions, emphasized the possibility of compactifying  space by suitable identifications of points. 

Lema\^{\i}tre fully shared 
the common belief in the closure of space. In a talk  
given at the Institut Catholique de Paris (Lema\^{\i}tre, 1978),
the Belgian physicist expressed his view that Riemannian geometry
 ``dissipated the nightmare of infinite space".  His two 
major cosmological models (the non--singular, ``Eddington-Lema\^{\i}tre" model, 1927,
 and the singular,``hesitating universe" model, 1931) assumed positive space curvature. 
Thus Lema\^{\i}tre thoroughly discussed the possibility of 
elliptical space, that he preferred to the spherical one. 
Later, Lema\^{\i}tre (1958) also noticed the possibility of 
hyperbolic spaces as well as  Euclidean spaces with finite volumes for describing 
the physical universe.

Such fruitful ideas of cosmic topology remained widely ignored by 
the main stream of big bang cosmology. Perhaps the Einstein-de Sitter model 
(1932), which assumed Euclidean space and eluded the 
topological question, had a  negative influence on the 
development of the field.  Almost all subsequent textbooks and 
monographies on relativistic cosmology assumed that the global 
structure of the universe was either the finite hypersphere, or the 
infinite Euclidean space, or the infinite hyperbolic space, without 
mentioning at all the topological indeterminacy.  As a consequence, some 
 confusion settled down about the real meaning of the terms ``open" and 
``closed" used to characterize the F--L  solutions, even in recent specialized 
articles (e.g. White and Scott, 1996). Wheras they apply correctly to
 time evolution (open models stand for ever--expanding universes, closed 
 models stand for expanding--contracting solutions), they do not 
 properly describe the extension of 
space (open for infinite, closed for finite). Nevertheless it is 
still frequent to read that the (closed) spherical 
model has a finite volume whereas the (open) Euclidean and hyperbolic models 
have infinite volumes. The 
correspondance is true only in the very special case of a 
simply--connected topology {\it and} a zero cosmological constant.  
According to Friedmann's original remark, in order to know if a
space is finite or infinite, it is not sufficient to determine the 
sign of its spatial curvature, or equivalently in a cosmological 
context to measure the ratio 
of the average density to the critical value~: additional assumptions 
are necessary - those arising from topology, precisely.

Until 1995, investigations in cosmic topology were rather scarce (see  
references in LaLu95). 
From an epistemological point of view, it 
seems that the prejudice in favour of simply--connected (rather 
than multi--connected) spaces was of the same kind as the prejudice 
 in favour of static (rather than dynamical) 
cosmologies during the 1920's.  At a first 
glance, an ``economy principle" (often useful in physical modelling) could 
be invoked to preferably select the simply--connected topologies.  However, 
on one hand, new approaches of spacetime, such as quantum cosmology, 
suggest that the smallest closed hyperbolic manifolds are favored 
(Cornish, Gibbons \& Weeks, 1998), thus providing a new paradigm for what is the 
``simplest" manifold. On 
the other hand, present astronomical data indicate that the average 
density of the observable universe is less than the critical value 
($\Omega = 0.3 - 0.4$), thus suggesting that we live in a negatively--curved 
F--L universe. Putting together these two requisites, 
cosmologists must face the fact that a negatively--curved space with 
a finite volume is necessarily multi--connected.  

\section {New developments} 

 In the last decades, much effort in observational and theoretical cosmology  
has been directed towards determining the curvature of 
the universe. The problem of topology of spacetime was generally 
ignored within the framework of classical relativistic cosmology. It 
began to be seriously discussed in quantum gravity for various reasons : the spontaneous birth 
of the universe from quantum vacuum  requires the universe 
to have compact spacelike hypersurfaces, and the closure of space is
 a necessary condition to render tractable the integrals of quantum  
gravity (Atkatz \& Pagels, 1982). 
However, the topology of spacetime also enters in a 
fundamental way in classical general relativity.  Many 
cosmologists were surprisingly unaware of how topology and cosmology 
could fit together to provide new insights in universe models.  Aimed to create a new 
interest in the field of cosmic topology, the 
extensive review by LaLu95 stressed on what multi--connectedness of the 
Universe would mean and on its observational consequences. However two different 
papers (Stevens et al, 1993 ; de Oliveira Costa \& Smoot, 1995) declared that the small 
universe idea was ``dead" ; in fact, drawing general conclusions from few examples mostly taken 
in the Euclidean case, they did not take into account the most 
interesting spaces for realistic universe models, namely the compact hyperbolic 
manifolds, which require a quite different treatment (LaLu95, Cornish et 
al., 1997a). Ironically enough, a worldwide interest for the 
subject has flourished  since 1995, both from an observational point of view and from a 
theoretical one : approximately the same amount of papers in cosmic topology have been 
published within the last 3 years as in the previous 80 years ! 
Interesting progress has been achieved in mathematics as well as in 
cosmology. I briefly summarizes below some of these advances.

\subsection {Mathematical advances} 

Just remember here that the method for classifying the admissible topologies of a manifold 
is to 
determine the universal covering space and the fundamental polyhedron, 
and to calculate the holonomy group acting on the polyhedron.  
Particularly important cases for application to  cosmology 
are the locally homogeneous and isotropic 3 -- dimensional 
Riemannian manifolds, {\it i.e.} admitting one of the three geometries 
of constant curvature. The correspondance between the local homogeneous 3--geometries 
in Thurston's sense and the Bianchi--Kantowski--Sachs classification
 of homogeneous cosmological models has been fully clarified (Rainer, 1996).
 
 Any space of constant curvature $M$ can  
be expressed as a quotient $M = \tilde{M}/\Gamma$, where the 
universal covering space $\tilde{M}$ is either the Euclidean space ${\bf R}^3$ if
 $K = 0$, the hypersphere ${\bf S}^3$ if $K > 0$ or the hyperbolic 
3-space ${\bf H} ^3$ if $K < 0$, and $\Gamma$ is a discrete subgroup of 
isometries without fixed point of $\tilde{M}$. 

Classification of 
Euclidean and spherical spaceforms being achieved (see Wolf, 1984), only the case 
of three-dimensional homogeneous hyperbolic  manifolds was recently investigated. 
The first thing to retain is that the 2--dimensional case does 
not give a good intuition of what can happen in higher dimensions. The Mostow theorem 
illustrates an essential difference between 2-dimensional hyperbolic 
geometry and higher dimensions : while a surface of genus $\ge 2$ 
supports uncountably many non equivalent hyperbolic structures, for $n 
\ge 3$ a connected oriented n--dimensional manifold supports {\it at 
most one} hyperbolic structure.  It follows that geometric invariants 
such as the volume, or the lengths of closed geodesics are also 
topological invariants, and the  curvature radius is a characteristic length scale
 for topology. It is the reason why the tentative 
classification of CHMs is based on increasing 
volumes.

 Each CHM topology has a specific volume measured in curvature radius units.
The absolute lower bound for the volume of CHMs, given previously as $V_{min}
= 0.000082$, has been raised to $V_{min} = 0.16668$ (Gabai et al., 1996). Fortunately it has little effect on cosmological applications.
The reason is that the true lower bound is almost surely 0.942707 
(Weeks, 1998), corresponding to the smallest CHM that is presently known  
(Weeks, 1985; Matveev and Fomenko, 1988). The new $V_{min}$ bound represents 
an improvement in the techniques
of the proof, not an increase in the expected size of the smallest
hyperbolic manifold. 

The WFM manifold leaves room for many topological lens effects,
since the volume of the observable universe is about 200 times larger than the volume of 
WFM space (Costa \& Fagundes, 1998). Indeed, many CHMs have 
geodesics shorter than the curvature radius, leaving room to fit a great many
copies of a fundamental polyhedron within the horizon radius, even
for manifolds of volume $\sim$ 10.
The publicly available program {\sc SnapPea} (Weeks) is 
specially useful to unveil the rich structure of CHMs. 
Table I summarizes some of the results ($r_{-}$ is the radius of the largest 
sphere in the covering space  which can be inscribed in the fundamental polyhedron,
$r_{+}$ is the radius of the smallest 
sphere in the covering space in which the fundamental polyhedron can 
be inscribed, $l_{min}$ is the 
length of the shortest geodesic).

\begin{table}
\caption{{\bf The smallest known CHMs}}
\begin{center}
\begin{tabular}{ccccc}
\hline
Name & Volume& $r_-$ & $r_+$ & $l_{min}$ \\
\hline
WMF   & 0.9427 & 0.5192 & 0.7525 & 0.5846 \\
Thurston   & 0.9814 & 0.5354 & 0.7485 & 0.5780 \\
s556(-1,1) & 1.0156 & 0.5276 & 0.7518 & 0.8317 \\
m006(-1,2) & 1.2637 & 0.5502 & 0.8373 & 0.5750 \\
m188(-1,1) & 1.2845 & 0.5335 & 0.9002 & 0.4804 \\
v2030(1,1) & 1.3956 & 0.5483 & 1.0361 & 0.3662 \\
m015(4,1)  & 1.4124 & 0.5584 & 0.8941 & 0.7942 \\
s718(1,1)  & 2.2726 & 0.6837 & 0.9692 & 0.3392 \\
m120(-6,1) & 3.1411 & 0.7269 & 1.2252 & 0. 3140 \\
s654(-3,1) & 4.0855 & 0.7834 & 1.1918 & 0. 3118  \\
v2833(2,3) & 5.0629 & 0.7967 & 1.3322 & 0. 4860 \\
v3509(4,3) & 6.2392 & 0.9050 & 1.3013 & 0.3458
\end{tabular}
\end{center}
\end{table}

\subsection{Cosmological advances}

The global topology of the universe can be tested by studying the 
3--D distribution of discrete sources and the 2--D fluctuations in the 
Cosmic Microwave Background (CMB). The methods are all based on 
the search for ``ghost images" predicted 
by Friedmann (1924), namely topological images of a same celestial 
object such as a galaxy, a cluster or a spot in the CMB (the term ``ghost" can 
lead to a confusion in the sense that all the images are on the same foot of reality). 
Such topological images can appear in a multi-connected  space a characteristic length 
scale of which is smaller 
than the size of the 
observable space, because light emitted by a distant source can reach 
an observer along several null geodesics. In the 1970's, systematic 
2--D observations of galaxies,
 undertaken at the 6--m Zelentchuk telescope under the 
supervision of Schvartsman, allowed to fix lower 
limits to the size of physical space as $500 h^{-1}Mpc$ (see LaLu95 
and references herein). 
A new observational test based on 3--D analysis of clusters 
separations, the so--called ``cosmic 
crystallographic method", has been  proposed by Lehoucq, 
Lachi\`eze-Rey \& Luminet (1996), and 
further discussed in the literature (Fagundes \& Gausmann, 1997; 
Roukema \& Blanloeil, 1998). Other 3--D methods, using special quasars 
configurations (Roukema, 1996) or X-ray clusters (Roukema \& Edge, 
1997), have also been devised, see Roukema (1998) for a summary. 
However, the poorness of 3--D data presently limits the power of such 
methods. 

Some authors (de Oliveira-Costa et al., 1996), still believing that 
an inflationary scenario necessarily leads 
to an Einstein-de Sitter universe, looked for constraints on topology 
with the CMB by investigating the compact Euclidean 
manifolds (CEM) only. They found that toroidal 
universe with rectangular cells (the simplest CEM, described as $E_{1}$ 
in LaLu95's classification), with cell size smaller than $3000 h^{-1} 
Mpc$ for a scale--invariant power spectrum, were ruled out as 
``interesting cosmological models". However, as shown by Fagundes \& 
Gausmann (1997), CEMs remain physically meaningful 
even if the size of their spatial sections is of the same order of 
magnitude as the radius of the observable horizon. Using the method of 
cosmic crystallography (Lehoucq et al., 1996) they performed 
simulations showing sharp peaks 
in the distribution of distances between topological images.

In any case, CHMs  appear today as the most promising specimens for cosmology, both from 
theoretical and observational grounds. 
Topological signature using the ``circles in the sky" method 
(Cornish et al., 1996b) is 
difficult to detect in COBE data, but it could be possible with the 
future MAP and/or PLANCK data. 

In fact, full cosmological calculations in CHMs (e.g. simulations of CMB fluctuations, or 
possible Casimir-like effects in the early universe) are 
difficult; they require calculations of eigenmodes of the Laplace
operator acting on the compact manifold. The problem is not solved.
 Cornish \& Turok (1998) recenly suggested a method working in 2--D, 
 but the 3--D case could reveal untractable. Compactness renders the calculations more difficult due to ``geodesic 
 mixing", namely chaotic behaviour of geodesic bundles (Cornish et 
 al., 1996a). Some authors (Levin et al, 1997, Cornish et al., 1997) were able to perform calculations
 in the ``horn topology", an open hyperbolic space introduced by Sokoloff and 
 Starobinsky (1976), but the essential job remains to be done.
 
 Another underdeveloped promising field is the interface between topology 
 and the early universe at high energy (although below the Planck 
 scale). Uzan and Peter (1997) showed that if space 
 is multi--connected on scales now smaller than the horizon size, the 
 topological defects  such as strings, domain walls, \ldots expected 
 from GUT to arise at phase transitions, were very unlikely to form.
 
In my opinion, a major breakthrough in the field of cosmic topology 
would be to relate the topological length scale $L_{T}$ 
with the 
 cosmological constant $\Lambda$. In an unified scheme
  with two fundamental lengths scales -- the Planck scale $l_{P}$ 
 and the inverse square root of the cosmological 
constant $\Lambda$, a consistent theory of 
quantum gravity should be able to predict the most probable value of  
$L_{T}$ in terms of $l_{P}$ and $\Lambda^{-1/2}$. Preliminary 
calculations in 2--D gravity models can be performed to test the idea.

\vskip .5cm 
\noindent 
{\Large\bf References} 
\vskip .5cm

\noindent  Atkatz, D., Pagels, H., 1982, Phys. Rv. D {\bf 25}, 2065

\noindent Bieberbach, L., 1911, {\it \"Uber die Bewegungsgruppen der Euklidischen 
Raume}, Mathematische Annalen, {\bf 70},297. 1912, Ibid, {\bf 72}, 400. 

\noindent Cornish, N.J., Spergel, 
D.N., Starkman, G.D., 1996a, Phys. Rev.Lett.{\bf 77}, 215.

\noindent Cornish, N.J., Spergel, D.N., Starkman, G.D., 1996b, gr-qc/9602039

\noindent Cornish, N.J., Spergel, D.N., Starkman, G.D., 1997, 
astro-ph/9708225

\noindent Cornish, N.J., \& Turok, N.G., 1998, {\it Ringing the 
eigenmodes from compact manifolds}, preprint gr-qc/9802066 (to appear in Class. Quant. Grav.)

\noindent Cornish, N.J., Gibbons, G., Weeks, J.R., 1998 (in 
preparation)

\noindent  Costa, S., Fagundes, H., 1998, {\it Birth of a Closed Universe of Negative 
Spatial Curvature}, preprint  gr-qc/9801066.

\noindent de Oliveira-Costa A., Smoot G., 1995, Astrophys. J. 
{\bf 448}, 477.

\noindent de Oliveira-Costa A., Smoot G., Starobinsky A., 1996, Astrophys. J. 
{\bf 468}, 457.

\noindent de~Sitter, W., 1917, MNRAS, {\bf 78}, 3.

\noindent Einstein, A., 1917, {\it Kosmologische Betrachtungen zur allgemeinen
 Relativit\"atstheorie}, Preussische Akademie der Wissenschaften,
  Sitzungsberichte, pp. 142--152

\noindent Einstein, A., 1918,  Postcard to Hermann Weyl, June, from
Einstein Archives, Princeton (freely translated from German by J.-P. 
Luminet).

\noindent Einstein, A., 1919,  Postcard to Kelix Klein, April 16th, ibid. 

\noindent Einstein, A., 1934, {\it Essays in Science} (New York : 
Philosophical Library) p. 52.

\noindent Einstein, A., de Sitter W., 1932, Proc. Nat. Acad. Sci {\bf 
18}, 213.

\noindent Fagundes, H. \& Gausmann, E., 1997, {\it On Closed 
Einstein-de Sitter Universes}, astro-ph/9704259.

\noindent Feodoroff, E., 1885, {\it Symmetrie der 
regelmassigen Systeme der Figuren}, Russian journal for crystallography 
and mineralogy, St Petersburg, {\bf  21}. 

\noindent Friedmann, A., 1922, Zeitschrift f\"ur Physik, {\bf 10}, 37.

\noindent Friedmann, A.,  1923, {\it Mir kak prostranstvo i 
vremya} (The Universe as Space and Time), Leningrad, Akademiya (French translation 
in {\it Essais de Cosmologie}, 
Le Seuil/Sources du Savoir, Paris, 1997.)

\noindent Friedmann, A.,1924, Zeitschrift f\"ur Physik, {\bf 21(5)}, 326.

\noindent Gabai, D. Meyerhoff, R., Thurston, N., 1996, {\it Homotopy Hyperbolic
3-Manifolds Are Hyperbolic},  preprint available  at 
http://www.msri.org/MSRI-preprints/online/1996-058.html

\noindent  Harrison, E., 1987, {\it Darkness at night}, Harvard University Press.

\noindent Klein, F., 1890, {\it Zur nicht-euklidischen 
Geometrie}, Mathematisches Annalen, {\bf 37}, 544.

\noindent Killing, W., 1891,
{\it \"Uber die Clifford-Kleinschen Raumformen}, Mathematisches 
Annalen, {\bf 39}, 257. 
 
\noindent Lachi\`eze-Rey, M. \& Luminet, J.-P., 1995 (LaLu95), 
Phys.Rep. {\bf 254}, 136.

\noindent Lehoucq, R., Lachi\`eze-Rey, M. \& Luminet, J.-P., 1996, Astron. Astrophys.,  
{\bf 313}, 339.

\noindent Lema\^{\i}tre, G., 1927,  Ann. Soc. Sci. Bruxelles, 
 ser. A, {\bf 47}, 29.

\noindent Lema\^{\i}tre, G., 1931, MNRAS {\bf 90}, 490.

\noindent Lema\^{\i}tre, G., 1958, in {\it La 
Structure et l'Evolution de l'Univers}, Onzi\`eme Conseil de Physique 
Solvay, ed.  Stoops, R., (Brussels: Stoops), p.1

\noindent Lema\^{\i}tre, G., 1978 (posth.) in {\it L'Univers, 
probl\`eme accessible \`a la science humaine}, 
Revue d'Histoire Scientifique, {\bf 31}, pp.  345-359.

\noindent Levin J., Barrow J.D., Bunn E.F., Silk J., 1997, Phys. rev. 
Lett. {\bf 79}, 974.

\noindent Luminet, J.-P, 1994 {\it L'infini 
dans la cosmologie relativiste}, in {\it Histoire et Actualit\'e de la 
Cosmologie, volume II}, \'ed.  F. De Gandt \& C. Vilain, Observatoire de 
Paris, pp.  85-98.

\noindent Luminet, J.-P., 1997 {\it L'invention du big bang}, 
suivi de {\it A. Friedmann, G. Lema\^{\i}tre : Essais de Cosmologie} (Le 
Seuil/Sources du Savoir, Paris).
 
\noindent Luminet, J.-P. \& Lachi\`eze-Rey, M., 1994, {\it La physique 
et l'infini} (Flammarion/Dominos, Paris).

\noindent Matveev S.V., Fomenko A.T., 1988, Russian Math.  
Surveys {\bf 43}, 3. 

\noindent Novacki, W., 1934, {\it Die euklidishen, dreidimensionalen, 
geschlossenen und offenen 
Raumformen}, Commentarii Mathematici Helvetici, {\bf 7}, 81.  

\noindent Roukema, B.F., 1996, MNRAS, {\bf 283}, 1147. 

\noindent Roukema, B.F., 1998, these proceedings (astro-ph/9801225).

\noindent Roukema, B.F., Edge, A.C., 1997, MNRAS {\bf 292}, 105.  

\noindent Roukema, B.F., Blanloeil V., 1998, {\it Three--dimensional 
Topology-Independent Methods to Look for Global Topology}, 
astro-ph/9802083 (to appear in Class. Quant. Grav.)

\noindent Rainer M., 1996, {\it Classifying spaces for homogeneous manifolds 
and their related Lie isometric deformations},
preprint gr-qc/9602059.

\noindent  Sokoloff D.D., Starobinsky A. A., 1976, Sov. Astron. {\bf 
19}, 629.

\noindent Stevens D., Scott D., Silk J. : Phys.  Rev.Lett.  {\bf 71}, 20 (1993)

\noindent Thurston, W.P. : {\it The geometry and 
topology of three manifolds}, (Princeton Lecture Notes, 1979);
ibid.,{\it Three-dimensional Geometry and Topology}, ed. S. Levy  
(Princeton University Press, 1997)

\noindent Uzan, J.-P., Peter P., 1997, Phys.Lett. B {\bf 406}, 20

\noindent Weeks, J., 1985, PhD thesis, Princeton University.

\noindent Weeks, J., {\it Snap Pea} : a computer program for creating and studying
 hyperbolic 3--manifolds, available at http://www.geom.umn.edu/software.

\noindent Weeks, J., 1998, private communication.

\noindent White M., Scott D., 1996, Astrophys. J. {\bf 459}, 415.

\noindent Wolf, J.A., 1960, Comptes rendus de l'Acad\'emie des Sciences de Paris, vol.  
{\bf 250}, 3443.

\noindent Wolf, J., 1984, {\it Spaces of constant curvature}, Fifth Edition, 
Publish or Perish Inc, Wilmington (USA).

\end{document}